# Breaking Bottlenecks in Solid Electrolyte Discovery with Large Artificial Intelligence Models

Eric Jianfeng Cheng[1,*], Min Hong[1], Zhiquan Zeng[2], Chuanyu Liu[3], Qian Wang[1], Melissa Jane Meadowcroft[4], Vlad Badilita[5], Carlos Miguel Costa[6], Senentxu Lanceros-Méndez[7], Ying Li[1], Chenyao Ma[8], Weibo Gong[9], Surendra Martha[10], Aloysius Soon[11], Piao Ma[8,9], Di Zhang[12], Teguh Ariyanto[13], Sudaryanto[14], Hiroshi Kakinuma[1], Jie Zhao[15], Jiayu Peng[3,*], Pengfei Ou[2,*], Jun Lu[16], Shin-ichi Orimo[1,17], Hao Li[1,*]

[1] Advanced Institute for Materials Research (WPI-AIMR), Tohoku University, Sendai 980-8577, Japan

[2] Department of Chemistry, National University of Singapore, Singapore

[3] Department of Materials Design and Innovation, University at Buffalo, Buffalo, NY 14260, USA

[4] London School of Economics, London WC2A 2AE, United Kingdom

[5] Institute of Microstructure Technology, Karlsruhe Institute of Technology, Eggenstein-Leopoldshafen 76344, Germany

[6] Physics Centre of Minho and Porto Universities (CF-UM-UP), University of Minho, Braga 4710-057, Portugal

[7] Basque Center for Materials, Applications and Nanostructures, UPV/EHU Science Park, Leioa 48940, Spain

[8] Suzhou MatSource Technology Co., Ltd., Suzhou 215000, China.

[9] Gusu Laboratory of Materials, Suzhou 215000, China

[10] Department of Chemistry, Indian Institute of Technology Hyderabad, Telangana 502284, India

[11] Department of Integrated Science and Engineering, Yonsei University, Incheon 21983, Republic of Korea

[12] The Frontier Research Institute for Interdisciplinary Sciences (FRIS), Tohoku University,

Sendai 980-8577, Japan

[13] Department of Chemical Engineering, Faculty of Engineering, Universitas Gadjah Mada 55281 Yogyakarta, Indonesia

[14] Research Center for Advanced Materials, National Research and Innovation Agency (BRIN), Tangerang Selatan 15311, Indonesia

[15] Department of Materials Science, Fudan University, Shanghai 200433, China

[16] College of Chemical and Biological Engineering, Zhejiang University, Hangzhou 310058, China

[17] Institute for Materials Research, Tohoku University, Sendai 980-8577, Japan

**Correspondence:** Eric Jianfeng Cheng (ericonium@tohoku.ac.jp), Jiayu Peng (jypeng@buffalo.edu), Pengfei Ou (pengf.ou@nus.edu.sg), Hao Li (li.hao.b8@tohoku.ac.jp)

**Abstract**: Solid electrolytes (SEs) are central to next-generation metal batteries, yet their discovery remains constrained by fragmented data, limited transferability of simulations, and slow experimental iteration. Unlike catalysis, where surface reactivity dominates, SEs require simultaneous optimization of bulk ion transport, defect chemistry, mechanical integrity, and interfacial stability. Here, we outline a framework for autonomous SE discovery enabled by large artificial intelligence (AI) models, including machine learning interatomic potentials (MLIPs) and large language models (LLMs). We discuss the evolution from static materials databases to dynamic, self-updating knowledge systems, the role of MLIPs in bridging density functional theory (DFT) and long-timescale ion migration, and the emergence of LLMs as engines for literature mining, hypothesis generation, and scientific reasoning. We further describe a closed-loop architecture integrating AI-driven candidate design, multiscale simulation, uncertainty-aware selection, and experimental validation. Such systems shift SE research from intuition-guided exploration to data-informed, self-improving cycles. We conclude by highlighting challenges in data standardization, interfacial complexity, and reproducibility, and we propose design principles for building autonomous laboratories for solid-state battery materials.

**Outlines**:

1. **Introduction: Why SEs need autonomous discovery**
- Coupled transport-mechanics-interface challenges
- Limits of traditional trial-and-error
- Lessons learned from AI in catalysis

2. **Databases for SEs**
- Existing computational and experimental repositories
- Data inconsistency and reporting challenges
- Dynamic databases and knowledge graphs

3. Machine learning interatomic potentials (MLIPs)
- MLIPs as a bridge between first-principles accuracy and battery-relevant scales
- Defect, grain boundary, and interface modeling
- High-throughput screening with physics fidelity

4. **Large Language Models (LLMs) in SE Research**
- Literature extraction and knowledge synthesis
- Hypothesis generation and design suggestions
- Multi-agent AI architecture

5. **Closed-Loop SE Discovery**
- From screening to feedback-driven discovery
- ML/HPC-guided discovery of halide SEs

- LLM-guided mining of MOF SEs
- AI-assisted 3D printing for SE manufacturing

**6. Outlook: Toward autonomous SE laboratories**

- Standardization and FAIR data
- Interface-aware AI
- Autonomous SE discovery beyond lithium
- Digital battery ecosystem

## 1. Introduction

Solid electrolytes (SEs) sit at the center of one of the most important unresolved problems in electrochemical energy storage: how to build batteries that are simultaneously safer, more energy dense, longer lasting, capable of faster charging, and scalable beyond today's liquid-electrolyte lithium-ion technology.[1–4] By replacing flammable organic liquid electrolytes with SEs, all-solid-state batteries (ASSBs) are expected to enable metallic Li or Na anodes, improve thermal and mechanical safety, and support compact cell configurations such as bipolar stacking.[5–9] Yet the field has learned a hard lesson: a high ionic conductivity alone does not make a practical SE. The central challenge is therefore not merely to identify fast ion conductors, but to design SEs that simultaneously combine high ionic conductivity, electronic insulation, electrochemical stability, mechanical robustness, processability, and durable interfaces with both cathodes and metal anodes.[10–13] Ionic transport depends on defect chemistry, lattice dynamics, local disorder, and migration-network topology; mechanical robustness is coupled to grain boundaries, porosity, and processing history; interfacial stability is controlled by chemical reactivity, space-charge effects, contact mechanics, and evolving decomposition products.[14–17] Consequently, SE performance emerges from a coupled transport-mechanics-interface landscape rather than from a single intrinsic descriptor. This complexity makes SE discovery particularly challenging and limits the efficiency of purely empirical optimization. This high-dimensional, interdependent, and experimentally costly design space is precisely why SE discovery requires autonomous workflows that can integrate data, simulation, experiment, and feedback-driven learning.

**1.1 Coupled transport–mechanics–interface challenges**

The historical evolution of SEs illustrates this trade-off-driven landscape. Oxides, sulfides, halides, polymers, hydrides, and composite electrolytes have each advanced specific aspects of SE performance, yet none has simultaneously resolved the coupled requirements of transport, stability, processability, and interfacial compatibility.[18–26] Oxide electrolytes such as garnets and NASICON-type materials offer attractive chemical stability and mechanical rigidity, but their brittleness and poor solid-solid contact remain significant barriers.[27] Sulfides can reach liquid-like ionic conductivities and form intimate interfaces under pressure, yet their moisture sensitivity and interfacial reactivity complicate manufacturing and long-term cycling. Halides have recently emerged as promising electrolytes with high oxidation potential, but their chemical diversity, air sensitivity, and interfacial behavior remain only partially understood.[28] Polymer and composite electrolytes offer processability and flexibility but often suffer from insufficient room-temperature conductivity or limited electrochemical stability.[29] More broadly, many SSB failures originate at electrode-electrolyte interfaces, where interfacial voids, chemo-mechanical coupling, and the continuous evolution of solid-electrolyte interphases during cycling can dominate over intrinsic bulk transport. No electrolyte family currently satisfies all practical criteria, and the trade-offs are not static; they change with composition, processing, electrode pairing, stack pressure, and cell architecture.

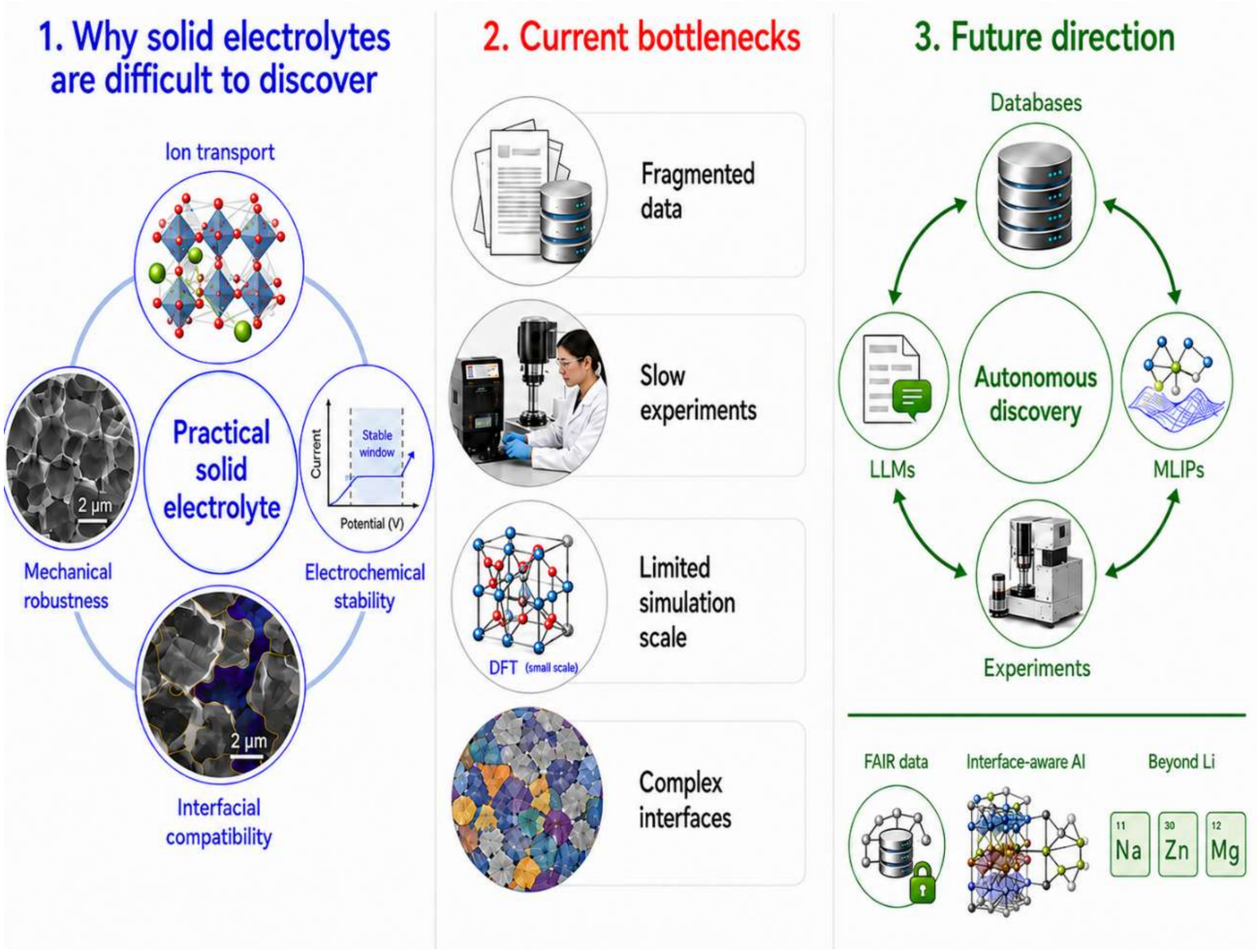

**Figure 1. Rationale and roadmap for autonomous SE discovery.** Practical SEs require coupled optimization of ion transport, stability, mechanics, and interfaces, while fragmented data, slow experiments, restricted simulation scales, and complex interfacial phenomena limit current progress. Integrating databases, MLIPs, LLMs, and experiments into autonomous loops offers a path from trial-and-error exploration toward data-driven discovery.

### 1.2 Limits of traditional trial-and-error

Traditional SE discovery has therefore relied on chemical intuition, incremental compositional substitution, high-temperature synthesis, impedance-based evaluation, and mechanistic interpretation after experimental observation. Recent Bayesian-optimization-guided synthesis of $Li_7SiPS_8$ further illustrates how data-driven experimental design can identify non-intuitive processing windows that substantially improve ionic conductivity while reducing synthesis temperature and time.[30] This approach has delivered important breakthroughs, but it cannot efficiently explore the vast compositional and structural space now facing the field. Density functional theory (DFT), *ab initio* molecular dynamics (AIMD), and high-throughput computations have provided powerful insights into phase stability, migration barriers, and electrochemical windows, but they remain limited by timescale, length scale, and the difficulty of representing realistic defects, grain boundaries, and interfaces. At the same time, experimental screening remains constrained by synthesis throughput, sensitivity to processing conditions, and incomplete reproducibility across laboratories.[31] These limitations are further amplified by small and heterogeneous data: experimental and computational records are scattered across papers and repositories, property values are reported under inconsistent reference conditions, and theoretical predictions often lack sufficient information on precursor availability, reaction pathways, and practical synthesizability. The result is a widening gap between the complexity of SE design and the tools traditionally used to navigate it.

**1.3 Lessons learned from AI in catalysis and materials discovery**

Artificial intelligence (AI) offers a route to narrow this gap, but its role in SE research must be framed carefully.[2,32–34] In catalysis and related energy-materials fields, AI has already shown value in extracting structure–property relationships, accelerating screening, and guiding high-throughput experimentation.[35,36] However, SEs pose a different kind of challenge. Catalytic performance is often governed by surface adsorption, reaction barriers, and active-site chemistry, whereas SE performance is governed by bulk ion transport, defect-mediated dynamics, microstructure, and buried interfaces that evolve during cycling.[37–40] Directly transferring AI strategies from catalysis or molecular design to SEs is therefore not viable. SE discovery requires models that can integrate atomistic transport, materials processing, electrochemical stability, and interface degradation within a unified and uncertainty-aware framework.

Machine learning interatomic potentials (MLIPs) represent one important step in this direction.[41–45] By learning potential-energy surfaces from first-principles data, MLIPs can extend atomistic simulations to larger systems and longer timescales than conventional *ab initio* methods, making them particularly powerful for studying ion migration, disorder effects, grain boundaries, and reactive interfaces. Their significance lies not simply in making simulations faster, but in making ion transport accessible under conditions closer to real materials. In SEs, conductivity arises from thermally activated hopping, collective ion motion, lattice fluctuations, disorder, defects, and finite-size effects; it cannot be fully captured by isolated migration barriers alone. MLIPs, therefore, provide a potential bridge between first-principles accuracy and the statistical, structural, and temporal scales required to understand battery-relevant ion transport.

LLMs add a different but complementary capability.[46–49] Unlike MLIPs, which address the scale problem in atomistic modeling, LLMs address the fragmentation problem in scientific knowledge. SE literature contains a large but heterogeneous body of information spanning synthesis conditions, conductivity values, activation energies, phase identities, electrode configurations, interfacial treatments, and cycling protocols. Much of this information is embedded in text, figures, tables, and supporting information rather than in machine-readable datasets. LLMs can assist in extracting, normalizing, and linking this knowledge, while also supporting hypothesis generation, literature-grounded reasoning, and multi-agent workflow orchestration.[50] Their most important role is not to replace physical models or experiments, but to connect data, simulation, and decision-making into coherent research loops.

This convergence points toward a discovery paradigm: autonomous SE discovery enabled by large AI models. In such a framework, databases provide traceable and reusable knowledge; MLIPs evaluate ion transport and interfacial dynamics with improved physical fidelity; LLMs extract literature knowledge and coordinate reasoning; and closed-loop experiments validate and refine predictions. The goal is not blind automation. Rather, it is to transform SE research from intuition-guided trial-and-error into a self-improving cycle in which hypotheses are generated, tested, corrected, and reused with increasing efficiency.[2] The quality of this cycle will depend as much on data reliability, provenance, and standardization as on algorithmic sophistication.

In this *Perspective*, we outline a framework for autonomous SE discovery enabled by large AI models. We first discuss how SE databases are evolving from static repositories into dynamic, context-preserving knowledge systems. We then examine how MLIPs bridge first-principles theory and long-timescale ion transport, with emphasis on disorder engineering, correlated hopping topology, and finite-size-aware transport statistics. Next, we analyze the emerging role of LLMs in SE research, including literature extraction, knowledge synthesis, validation-coupled reasoning, and multi-agent architectures. We further describe closed-loop discovery workflows that combine AI-driven candidate generation, multiscale simulation, uncertainty-aware selection, and experimental validation. Finally, we discuss the outlook toward autonomous SE laboratories, highlighting the need for Findable, Accessible, Interoperable, and Reusable (FAIR) data, interface-aware AI, discovery beyond lithium-based systems, and the "digital battery ecosystem". Through this *Perspective*, we argue that the next major advance in SE research may come not from a single material family, but from a new research infrastructure capable of uniting data, models, experiments, and human insights into an adaptive discovery engine.

## 2. Databases for SEs

### 2.1 Existing computational and experimental repositories

The rapid expansion of SE research has created a large body of experimental and computational data, but this information remains highly fragmented across journal articles, supporting information, crystallographic repositories, computational databases, and laboratory records.[51,52] At first glance, this may appear to be a conventional data-management problem. In SE research, however, the challenge is deeper because the key property of interest, ionic transport, is not a context-free material property. A reported conductivity value depends on temperature, conducting ion, sample density, processing history, electrode configuration, equivalent-circuit fitting, atmosphere, stack pressure, and whether the value is directly measured, interpolated, or extracted from Arrhenius analysis.[53] Unlike general materials repositories, SE databases must preserve measurement context, temperature dependence, and processing provenance because ion transport cannot be meaningfully compared without them.

General-purpose materials databases such as the Materials Project and OQMD have played an essential role in computational materials science by providing crystal structures, formation energies, phase stability, and related descriptors.[51,52] These repositories are invaluable for identifying candidate compounds and constructing thermodynamic or structural features. However, they were not designed to organize experimental ion-transport data in a form directly usable for SE discovery. In particular, they rarely contain temperature-dependent conductivity, activation-energy fitting details, electrode configurations, pellet density, sintering conditions, or impedance-analysis protocols. As a result, they provide the structural and thermodynamic foundation for SE informatics, but not a complete SE-native data environment.

Dedicated SE datasets and databases have therefore become necessary. Expert-curated conductivity datasets have demonstrated that literature-derived SE data can support ML analysis when values are carefully standardized and validated.[53] Similarly, SE-focused platforms such as the Dynamic Database of Solid Electrolytes (DDSE, recently renamed as the Digital Battery Platform, *DigBat*: www.digbat.org) and its later extensions provide a more specialized data structure for organizing conductivity, activation energy ($E_a$), conducting ion, material family, and literature source.[54,55] A key advancement of such platforms is that temperature-dependent conductivity values are preserved as multiple records rather than compressed into a single room-temperature value.

The more recent integrated SE data landscape further extends this concept. As summarized in Figure 2a, the current platform includes 3736 experimental SEs with temperature-dependent ionic-conductivity records and 852 computational SEs with $E_a$ data.[49] The experimental records cover inorganic, polymer, and gel polymer electrolytes, while the computational records include migration barriers and simulation-derived quantities. By linking experimental conductivity, $E_a$ , material composition, structure type, and computational descriptors under related material identities, such a platform begins to support cross-validation between experiment and computation.

**2.2 Dynamic databases and knowledge graphs**

Dynamic SE databases can, therefore, change how research questions are asked. Instead of asking only, “Which material has the highest conductivity?”, researchers can ask more mechanistic questions: Which chemical families show low activation barriers under comparable conditions? How does the presence of neutral molecules influence ion migration in hydride electrolytes? Which computational methods best reproduce experimental $E_a$? Which descriptors correlate with transport barriers across divalent systems? Which underexplored chemical spaces contain candidates similar to known high-performance SEs?[48,49]

The examples shown in Figure 2b–e illustrate this transition from data storage to data-enabled interpretation. Figure 2b compares temperature-dependent ionic conductivity for hydride SEs with and without neutral molecules, allowing transport trends to be discussed in relation to local molecular environments.[49] Figure 2c compares computational and experimental $E_a$, providing a basis for evaluating which simulation methods more reliably capture ion migration. Figure 2d correlates $E_a$ in hydride SEs with theoretical descriptors, showing how curated data can support interpretable regression and mechanism-oriented analysis. Figure 2e further extends the database concept to candidate discovery, where representation clustering and literature mining are used to identify promising metal-organic framework (MOF) SEs.[48]

LLMs can further enhance this transition by extracting information from text, tables, figures, and supporting information, and by assisting with literature searches, data normalization, and relationship construction.[56,57] However, LLM-assisted extraction does not remove the need for careful data standards. On the contrary, it makes standardization more important. If the original literature does not clearly define the plotted quantity, unit convention, fitting range, or measurement context, an LLM may extract a plausible but incorrect value.[56]

### 2.3 Data inconsistency and reporting challenges

Despite rapid progress, current SE databases still face several fundamental limitations. The first is data imbalance. Positive results are overrepresented because published literature preferentially reports successful syntheses, high conductivity, stable cycling, or optimized compositions. Failed syntheses, low-conductivity phases, unstable interfaces, poor densification, and unsuccessful processing windows are rarely reported in a structured form. This creates a serious problem for AI-driven discovery because active learning and autonomous laboratories require not only examples of what works, but also reliable information about what does not work.

The second limitation is missing metadata. Many literature reports provide conductivity values but omit key contextual details such as exact measurement temperature, pellet density, electrode type, stack pressure, atmosphere, impedance-fitting model, equivalent circuit, activation-energy fitting range, or whether the reported value is measured or extrapolated.[53] Processing information is also often incomplete, even though sintering temperature, dwell time, pressure, atmosphere, particle size, and post-treatment can strongly influence microstructure and transport.

The third and most subtle limitation is reporting ambiguity. Recent data-reliability analysis shows that conductivity reporting in SE literature can suffer from text–figure inconsistencies, ambiguous Arrhenius annotations, unit discrepancies, incomplete measurement context, and difficult-to-verify outlier values. These issues may appear minor in individual papers, but they become consequential when values are extracted into databases and reused for ML, where poorly contextualized outliers can bias model training and distort structure–property relationships.

These limitations indicate that the next generation of SE databases should not be judged by entry number alone. Their real value will depend on whether they preserve provenance, encode uncertainty, include negative or failed results, and link transport properties to synthesis, processing, microstructure, and measurement protocols. A conductivity value without context may be useful for human reading, but it is insufficient as an ML label. For autonomous SE discovery, databases must therefore evolve from passive repositories into context-preserving knowledge infrastructures that can support validation, model training, uncertainty-aware screening, and closed-loop decision-making. This requirement provides the foundation for the MLIP-, LLM-, and feedback-driven discovery frameworks discussed in the following sections.

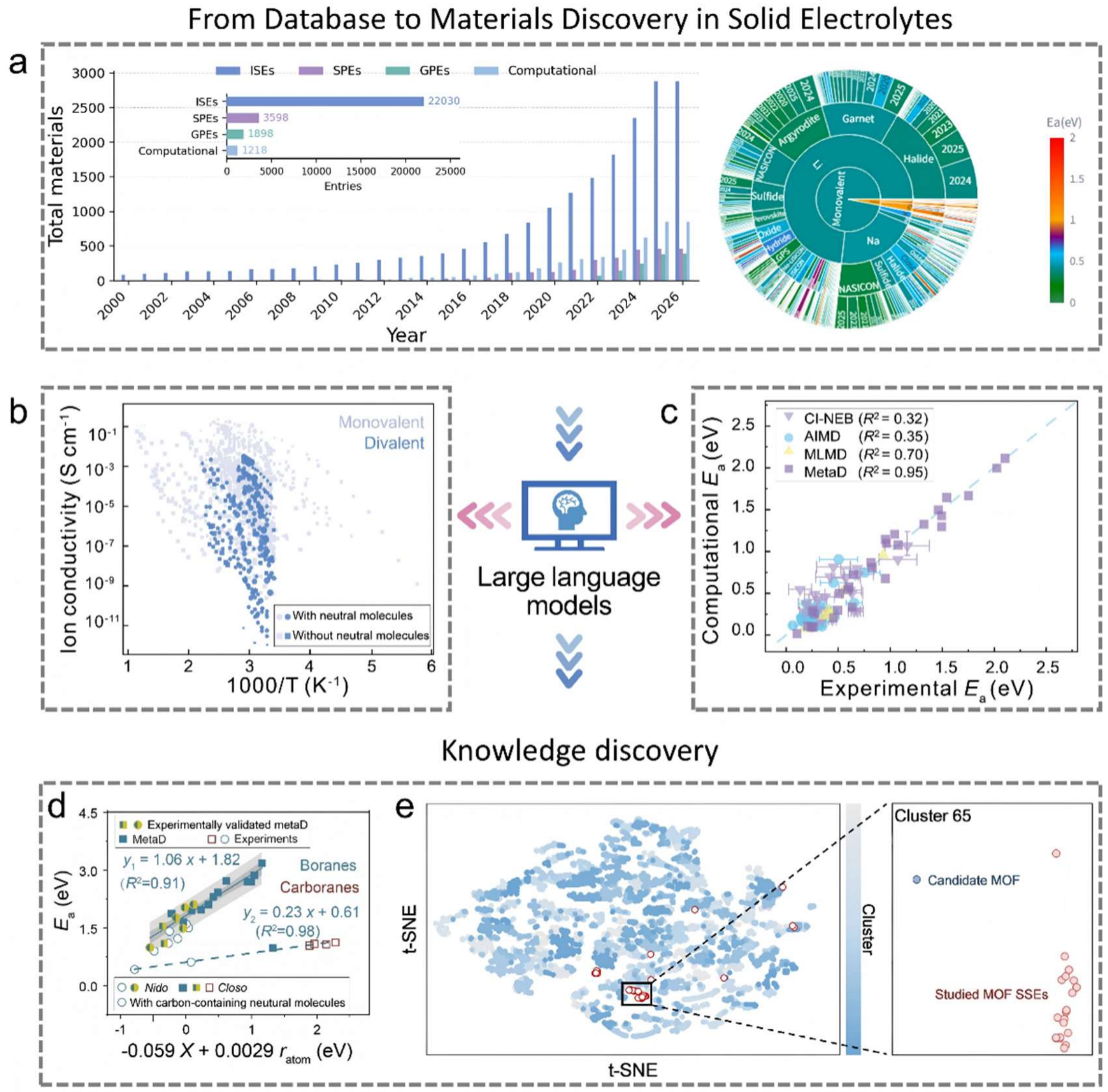

**Figure 2. Data integration-enabled materials discovery in SEs.** (**a**) Overview of the integrated SE data landscape, including 3736 experimental SEs with temperature-dependent ionic-conductivity records and 852 computational SEs with $E_a$ data. (**b**) Temperature-dependent conductivity for hydride SEs, with/without neutral molecules. (**c**) Comparison of computational and experimental activation energies for different methods. (**d**) Correlation of $E_a$ in hydride SEs with theoretical descriptors, including linear regression for divalent systems with neutral molecules. (**e**) Representation clustering for mining MOF SE candidates.

## 3. MLIPs for SEs

### 3.1. MLIPs as a bridge between first-principles accuracy and battery-relevant scales

MLIPs have emerged as a promising computational toolbox for understanding, designing, and optimizing SEs because they address a central mismatch between the scientific questions the battery research field asks and the physical scales that conventional atomistic methods can reach.[58] SEs must enable fast ion migration through a crystalline or partially disordered host, but their critical performance is also shaped by defects, grain boundaries, and chemically reactive interphases in close contact with electrodes.[59] DFT and AIMD remain the benchmark tools for atomistic fidelity, but they are too costly for the large supercells, long trajectories, and low-temperature statistics required to resolve rare hopping events, microstructural transport, and interfacial evolution. Classical force fields, by contrast, reach those length scales and timescales but often sacrifice accuracy and transferability, particularly when local chemistry, coordination, or disorder changes. MLIPs have therefore become valuable not simply because they are faster, but because they can potentially offer a route to near-DFT-level property prediction and materials screening at the time and length scales where solid-state battery phenomena unfold.

This detailed balance between physical fidelity and scale is why MLIPs should be viewed less as an alternative computational method and more as an enabling layer between first-principles theory and realistic battery physics (Figure 3a). Currently, two broad model categories coexist. System-specific MLIPs are trained for one chemistry or structural family and can be highly reliable provided that the training data already cover the relevant distortions,[60–63] defects, and nonequilibrium local environments. In contrast, universal or foundation MLIPs are pretrained across broader materials spaces and promise rapid deployment, but they are inherently more vulnerable when the computational task depends on activated hops, strained interfaces, or defect configurations that lie far from their original training distribution.[64,65] For SEs, this distinction matters because the most important observables are not merely energies and forces, but transport barriers, finite-size convergence, defect energetics, and interfacial stability.[66] A potential generator workflow for $Li_{10}GeP_2S_{12}$ (LGPS)-type superionic conductors showed how active-learning-driven MLIP training could be organized around diffusion statistics and activation energies rather than force fitting alone.[67] Also, an MLIP-based molecular dynamics (MD) benchmark for SEs highlighted that the effectiveness of an MLIP for ionic transport heavily depends on its training and validation.[68] Likewise, foundation-style MLIPs trained on out-of-equilibrium data can achieve broad transfer across LGPS-like, argyrodite, and mixed-composition sulfide electrolytes.[69] These key examples highlight that careful model construction and training are critical for building MLIPs targeted for SEs.

### 3.2. Bridging DFT and long-timescale ion transport

MLIPs' key contribution to SE research is bridging the gap between DFT and experimental ion transport. Ionic conductivity is a rare-event property: it depends on thermally activated hops, correlated motion, and sufficiently converged sampling over local environments, all of which are difficult to extract from short AIMD trajectories in modest supercells.[58] As a result, AIMD can yield mechanistic pictures that are correct generally but lack detail, particularly when high-temperature behavior is extrapolated to room temperature. MLIPs improve this situation by enabling nanosecond- to microsecond-scale MD while retaining a learned first-principles description of the potential-energy surface (PES).

This expanded window in timescale and length scale has already altered the mechanistic interpretation of solid-state ion conductors.[68] For example, long MLIP-powered MD simulations of LGPS showed that its non-Arrhenius behavior is tied to a Li-sublattice transition near 400 K, below which concerted hopping and cross-correlation decrease prominently (Figure 3).[70] Furthermore, a moment tensor potential (MTP) study showed that such MLIPs can narrow the gap between simulated and measured ionic conductivities.[71] More broadly, systematic comparisons demonstrated that MLIP-driven MD can reproduce distinct migration mechanisms, vibrational anharmonicity, and diffusion coefficients.[72] Together, these studies suggested that MLIPs are valuable not only for reproducing known diffusion trends more efficiently but also for enabling the retesting of mechanistic assumptions under experimentally determined sampling conditions, while increasingly revealing that collective ion migration is governed not only by isolated hopping barriers but also by the topology and persistence of correlated hopping pathways.

A second insight emerging from MLIP-enabled transport simulations is that disorder is often beneficial, but rarely in a simple monotonic way.[73] MLIP simulations of Cl-rich argyrodites, $Li_{6-x}PS_{5-x}Cl_{1+x}$ (LPSC), revealed a non-monotonic dependence of Li-ion conductivity on excess Cl and S/Cl site inversion, with high conductivity confined to a limited disorder window and a critical threshold beyond which long-range transport declines sharply. (Figure 3c).[74] In $Li_3PS_4$, another study showed that disorder could enhance room-temperature conductivity and introduced a softness descriptor.[75] A separate MLIP study of $Li_6PS_5Cl$ also reflected lattice-parameter effects and surface-related contributions that remain obscured in shorter simulation trajectories.[76] The broader implication is that MLIPs are beginning to better expose transport as a PES landscape shaped by local motifs, collective statistics, and subtle structural frustration. In turn, disorder is emerging not merely as a structural complication, but as a potentially controllable strategy for optimizing local transport landscapes.

This same logic has extended naturally to various families of chemistries of SEs.[77] An on-the-fly MLIP reproduced an $E_a$ close to experiment, and proposed a dynamic monkey-bar mechanism in which anion motion repeatedly opened and reconnected lithium pathways.[78] Moreover, a large-cell MLIP study of $Li_3YCl_6$ made finite-size effects impossible to ignore by showing that small AIMD cells could miss the temperature dependence and transport behavior needed for a realistic room-temperature conductivity estimate.[79] An iterative fine-tuning MLIP workflow for $Li_3YCl_{6-x}Br_x$ extended this idea by showing that a pretrained CHGNet model[80] can be further specialized to ordered and disordered halide structures, making conductivity trends across composition far more realistic.[81] These examples show MLIPs do more than speed up transport simulations; they help identify intrinsic transport features from artifacts of limited scale, thereby emphasizing that finite-size-aware transport statistics are essential for distinguishing intrinsic transport behavior from simulation artifacts.

Taken together, these studies suggest three emerging design principles for MLIP-guided SE discovery: (i) correlated hopping topology as a governing descriptor for collective ion migration beyond single-ion barriers, (ii) disorder engineering as a controllable strategy to optimize local transport landscapes, and (iii) finite-size-aware transport statistics to distinguish intrinsic transport behavior from simulation artifacts. Framed in this way, the significance of MLIPs is not only that they extend atomistic simulations to longer timescales, but that they enable ion transport in SEs to be interpreted using descriptors and statistical regimes that are more faithful to real materials behavior.

**3.3 Defect, grain-boundary, and interface modeling**

If long-timescale ion transport has established the initial credibility of MLIPs in this field, their next major contribution has been advancing atomistic modeling beyond idealized bulk structures. Practical SEs are polycrystalline, defect-rich, and chemically evolving materials. Their performance is limited more by the slowest kinetic bottleneck introduced by grain boundaries, secondary phases, interphases, or cation interdiffusion.[59] While DFT struggles with large, diverse simulation cells, MLIPs are especially valuable for battery performance that depends on mesoscale realism rooted in atomistic physics.

Grain boundaries exemplify this methodological shift. An MTP study of $Li_6PS_5Cl$ showed that MLIPs can resolve both bulk and grain-boundary diffusion mechanisms.[82] Moreover, a multiscale extension for $Li_6PS_5X$ (X = Cl, Br, I) coupled MLIP-derived MD diffusivities to continuum modeling and found that grain-size effects could reverse the role of grain boundaries based on the bulk transport regime (Figure 3d).[83] In other words, the question shifts from whether a grain boundary is blocking or beneficial alone, to how atomistic grain-boundary transport influences effective conductivity across microstructural length scales in electrolyte pellets.

Battery interfaces pose an even greater challenge test due to structural disorder, compositional gradients, and chemical reactivity.[84] A long-timescale MLIP study across sulfide, chloride, and oxide electrolytes with layered LCO identifies three transport blocking routes: interfacial reactions, Li-depleted regions, and cation interdiffusion.[85] In addition, an MLIP-assisted study showed that ionic interdiffusion can rapidly form a resistive interphase and that protective interlayers may suppress chemical damage while introducing new mechanical trade-offs.[86] In all-solid-state Li–S batteries, an MTP model shows that sulfide electrolytes are favorable against sulfur cathodes and that interfacial structure can reshape Li-migration barriers and buffer-layer design choices.[87] Extending MLIPs to interphase chemistry, a generative-structure-prediction and fine-tuned-CHGNet workflow suggested that amorphous $LiPO_2F_2$ may serve as a conductive interphase medium.[88] These results suggest MLIPs are shifting interface modeling from static assessments to dynamic problems involving changing composition, kinetics, and transport.

Electrochemical boundary conditions could represent the next frontier in that progression from realism to relevance. An on-the-fly MTP workflow for cathode coating materials likewise showed that MLIPs can extend MD into the moderate-conductivity regime relevant to interface coatings, where AIMD is often too short to yield reliable transport statistics.[89] A recent constant-potential MLIP framework with charge equilibration revealed that amorphous components within inorganic SE-interphase cause local lithium aggregation, triggering inhomogeneous deposition and nanoscale dendrite nucleation (Figure 3e).[90] Moreover, a related benchmark on interstitial Li diffusion in LiF using a pretrained MACE foundation model[91] highlights the growing relevance of transfer learning for interphase modeling.[92] This shift toward electrochemical realism also underscores a modeling issue: while short-ranged MLIPs can reasonably describe bulk Li transport, defect energetics and field-sensitive interfaces may require explicit treatment of long-range electrostatics.[93] What matters here is the specific metal electrodeposition mechanism and the methodological direction: MLIP frameworks shift from neutral bulk diffusion to electro-chemo-mechanical evolution under driving forces, aligning more with the failure modes that affect the success of SSBs. This shift toward electrochemical realism also underscores an important methodological consideration: while short-range MLIPs perform well for neutral bulk diffusion, defect energetics and field-driven processes at reactive interfaces often require explicit long-range electrostatics, as demonstrated in recent constant-potential MLIP simulations of Li-metal/SE interphases.[94]

### 3.4. High-throughput screening with physics fidelity

MLIPs would be deployed in high-throughput materials screening workflows once they could resolve transport and degradation at useful scales. Yet high-throughput screening must be approached cautiously, as the most dangerous failure mode is false screening confidence. A model that rapidly ranks thousands of candidates is only as useful as its fidelity in the activated, defective, and disordered regimes that control ionic transport.[58] The central question here is whether they can do so without drifting away from the relevant physics.

Current screening workflows have revealed several viable strategies.[95] A universal-MLIP workflow screened $7\times10^5$ Li-containing crystals, further filtered them using mechanical and electronic criteria, and reported about 130 candidate solid electrolytes, including halides such as $KLi_2ScBr_6$ and $Rb_2Li_3Br_5$, while explicitly noting that ion migration barriers were systematically underestimated and therefore required validation.[96] Another M3GNet-based workflow,[97] further refined through uncertainty estimation, screened 4,285 materials and identified eight promising candidates, demonstrating that universal-MLIP-based molecular dynamics can provide a practical front end for ion-conductivity prediction when confidence metrics are incorporated into the screening loop.[98] A complementary modeling strategy extracted potential-energy-landscape descriptors from a universal MLIP and efficiently ranked lithium materials, such that eight of the ten top candidates were later validated as room-temperature superionic conductors by first-principles calculations.[99] Similarly, a CHGNet-based high-throughput nudged elastic band workflow likewise identified Pnma-type $LiMgPO_4$ and $LiTiPO_5$ as promising ion conductors and extended the search to doped derivatives such as $Li_{0.5}Mg_{0.5}Al_{0.5}PO_4$ and $Li_{0.5}TiPO_{4.5}F_{0.5}$.[100] A related CHGNet prescreening study further showed that universal MLIPs can scan extremely massive NASICON and garnet design spaces for alkaline-stable compositions before a smaller subset is returned to DFT for refinement (Figure 3f).[101] Overall, these examples emphasize the promise of MLIPs in various high-throughput screening workflows for better SEs.

At the same time, physics fidelity must be treated as the primary currency of trust. In this regard, a recent benchmark based on the LiTraj dataset showed that MLIP models can distinguish high- from low-conductivity lithium conductors and predict migration barriers, thereby offering a task-specific standard for judging transport models rather than relying on generic force errors.[102] A broader assessment of SE research found that some MLIPs could reproduce energies, forces, elastic constants, and lithium diffusivities reasonably well, but remained chemistry- and task-dependent.[103,104] Complementarily, another analysis showed that several pretrained universal MLIPs systematically softened the PES in high-energy regions relevant to migration barriers, defects, and surfaces, although targeted fine-tuning can further correct much of the bias.[105] A further benchmark on surface energies also showed that out-of-domain prediction errors from MLIPs remain pronounced when local environments drift away from the model's training manifold, with obvious implications for grain boundaries and battery interfaces.[106]

In summary, the future for MLIPs in SE research will depend on more deliberate transport-aware workflows. SEs place unusually strict demands on atomistic MLIP models: they require accurate activated hops, chemically faithful defects, and dynamically evolving interfaces. Future methodological progress will therefore hinge on training sets that intentionally sample high-energy and defective environments, active-learning loops that target uncertainty, hybrid MLIP simulation schemes that handle electrostatics and variable charge, and validation protocols that also compare diffusivities, activation energies, and interfacial evolution against both DFT and experiments. In this sense, MLIPs have already started to shift SE modeling from small-cell, short-time, and largely idealized simulations into genuinely multiscale ones. They offer a route by which atomistic simulation can become genuinely more faithful to the heterogeneous, kinetic, and failure-prone reality of working SSBs.

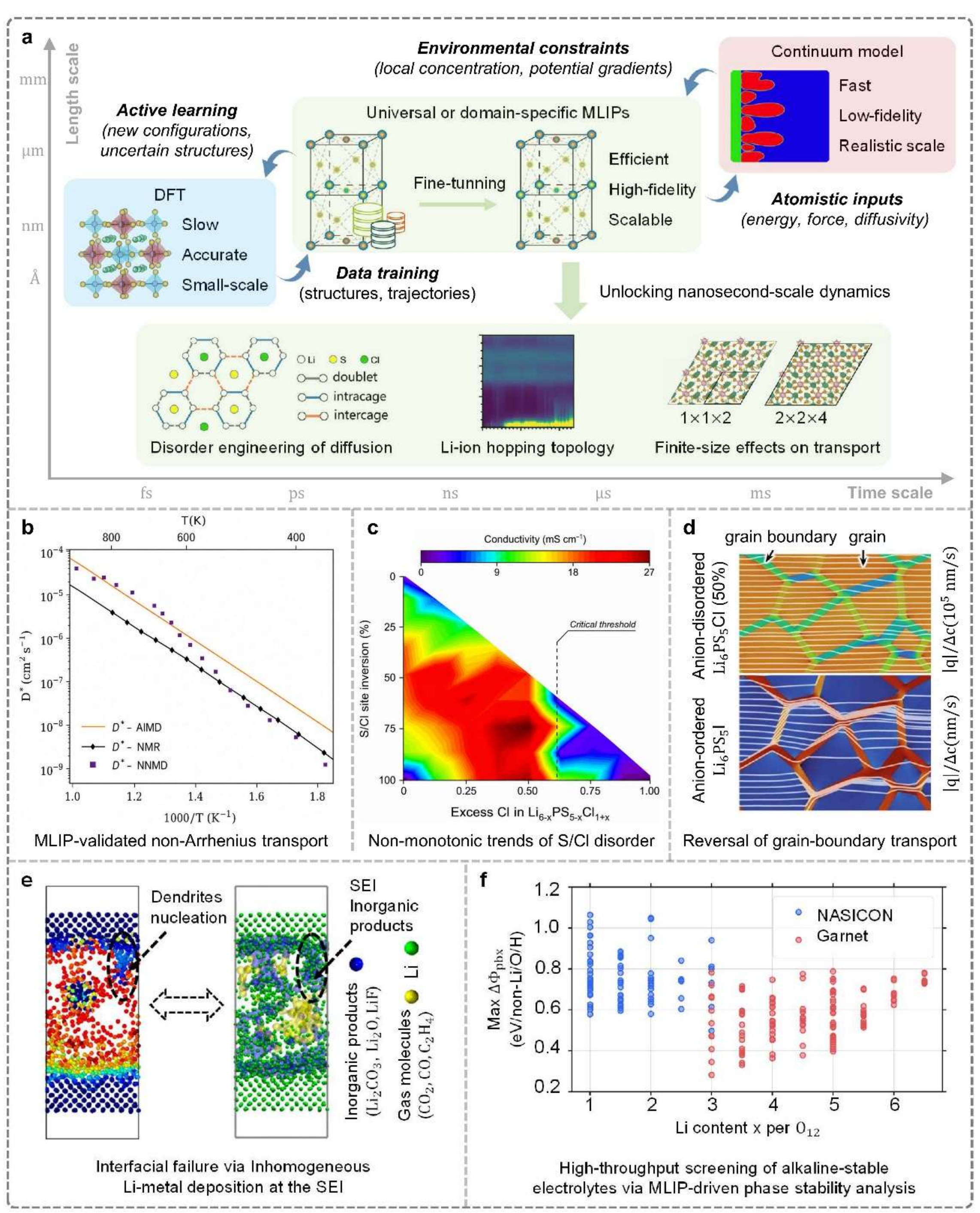

**Figure 3. Multi-scale modeling and validation of SEs. (a)** General hierarchical framework illustrating the spatiotemporal bridging by MLIPs. High-fidelity DFT data enables initial training,[107] while active learning loops identify uncertain configurations for continuous self-refinement. MLIPs achieve quantum accuracy at classical speeds,

supplying critical parameters, such as strain-dependent ionic diffusivities and interfacial energies, to continuum models,[108] which in turn define the macroscopic boundary conditions required to simulate complex phase evolution and failure modes. **(b)** Validation of MLIP-predicted transport kinetics against AIMD and experimental NMR data, demonstrating the model's ability to capture the non-Arrhenius behavior of Li-ion diffusion across a wide temperature range.[70] **(c)** MLIP-derived contour map showing the coupled effects of Cl incorporation and S/Cl site inversion on Li-ion conductivity in LPSC argyrodites.[74] Discovery of non-monotonic transport trends in sulfide SEs, where an intermediate S/Cl site disorder of 37.5–50% maximizes room-temperature ionic conductivity, highlighting the role of MLIPs in optimizing disordered structural environments.[109] **(d)** Continuum-scale dynamics of polycrystalline electrolytes, showing Li-ion flux maps where the role of grain boundaries shifts from blocking to enhancing transport based on the bulk transport regime, effectively bridging atomistic parameters to mesoscale morphology.[83] **(e)** Visualization of interfacial failure mechanisms at the Li-metal/SEI boundary; amorphous inorganic interphase components are shown to trigger local lithium aggregation and inhomogeneous deposition (dashed circles), preceding nanoscale dendrite nucleation.[90] **(f)** High-throughput screening of alkaline-stable electrolytes, identifying chemically robust NASICON and garnet frameworks capable of maintaining structural integrity under high-pH conditions through calculated phase stability metrics.[101]

## 4. LLMs in SE Research

### 4.1 LLMs as a knowledge and workflow layer

Beyond the atomistic simulation challenges discussed above, SE discovery is also limited by heterogeneous knowledge distributed across synthesis protocols, processing conditions, electrochemical measurements, and literature figures and tables. Unlike MLIPs, which primarily address the mismatch between first-principles accuracy and battery-relevant simulation scales, LLM-based systems address a different bottleneck: the fragmentation, inconsistency, and contextual ambiguity of scientific knowledge itself. In this setting, LLM-based systems are useful less as replacements for physical simulation than as a workflow layer that converts dispersed textual and graphical evidence into structured, queryable, and operational knowledge.[110,111] Earlier studies already showed that SE literature can be mined to recover processing parameters and ionic-transport data, while newer multimodal and domain-specific systems extend this logic toward knowledge synthesis, retrieval, and expert-facing research assistance.[112–115] These representative roles are summarized in Figure 4, which first provides an author-designed framework for extraction, reasoning, and execution systems (Figure 4a), and then illustrates representative examples of literature evidence curation (Figure 4b), SE-specific knowledge exploration (Figure 4c), and autonomous experimental execution (Figure 4d).

Related SE-focused studies have also begun to combine database construction, data mining, and LLM-assisted analysis to interrogate ion-migration trends and design workflows, as illustrated by the DDSE database-based viewpoints for Mg-ion SEs and by recent data-LLM-physics workflows for hydride electrolytes.[116–118] In this Perspective, we organize these contributions into three progressively integrated system classes: extraction systems, reasoning systems, and execution systems. This paper is also consistent with recent SE-focused discussions of AI agents, which describe SE-discovery systems in terms of perception, reasoning, action, and learning, while noting that most current implementations remain knowledge- and workflow-centered rather than fully autonomous closed-loop platforms.[119]

From the perspective of SE discovery, the importance of LLM-based systems lies not only in workflow automation, but also in the design principles they introduce for handling evidence. The first is evidence normalization: conductivity, migration barriers, synthesis conditions, and interfacial observations should not be compared across the literature until their provenance, measurement context, and reporting conventions have been aligned.[112–117,120] The second is validation-coupled reasoning: literature-grounded suggestions are useful only when they are linked to explicit downstream checks, such as screening criteria, mechanistic simulations, or synthesis constraints, rather than treated as free-form recommendations.[117,121,122] The third is execution-aware orchestration, in which expert knowledge, tool use, and iterative feedback are organized into machine-readable tasks that can be passed to simulation, database, or experimental modules.[123–126] In this sense, LLMs do not simply accelerate an existing workflow; they redefine SE discovery as a provenance-aware, validation-aware, and execution-aware design process. At present, however, this redefinition should be understood as a methodological direction rather than a mature autonomous capability, because most systems still depend on incomplete data, domain heuristics, and downstream human or physics-based verification.[127]

### 4.2 Literature extraction and knowledge synthesis

The first practical contribution of LLM-enabled systems in SE research is the conversion of dispersed literature into structured evidence rather than direct materials invention.[112,113] Early SE-specific studies showed that both synthesis conditions and ionic-conductivity records can be extracted systematically from the literature, establishing that process variables and property labels need not remain as isolated textual descriptions. This literature-to-database logic is also consistent with recent SE-focused data-driven viewpoints, which argue that broad and harmonized electrolyte datasets are necessary if design-relevant patterns are to be recognized beyond single reports.[116]

A major bottleneck is that many of the most informative SE data are not written explicitly in prose, but are buried in Arrhenius plots, impedance figures, and process schematics. Multimodal extraction systems address this limitation by jointly interpreting figures, tables, and text and converting them into provenance-aware structured outputs, which is especially relevant to SE literature because figure-level information often carries the data needed for later comparison and screening.[114] BatteryDataExtractor, a Python-based and battery-domain text-mining toolkit built on the BatteryBERT language model, provides a useful bridge between earlier rule-based extraction pipelines and transformer-based materials informatics. It represents a useful transition from earlier rule-based extraction pipelines to battery-aware transformer-based methods, while also highlighting a key limitation of current literature mining: many important data remain embedded in figures and tables and are therefore missed by text-only extraction workflows.[128]

To address the coupled challenges of figure–text synchronized extraction and consistency between material names and their associated properties,[128] Zhang et al. proposed the Descriptive Interpretation of Visual Expression, or DIVE, workflow. Rather than sending the whole PDF directly to a model, DIVE separates text and figures, and uses captions to locate key plots such as pressure-composition-temperature curves (PCT) and discharge curves.[129,130] The selected figure, caption, and nearby text are jointly read by a multimodal model. Prompt-guided conversion rewrites the curve information as descriptive text and reinserts it before final data extraction. This design helps preserve links among compositions, conditions, and properties when values are reported mainly in figures. In hydride papers, DIVE increased extraction accuracy and completeness by about 10 to 15% compared with commercial multimodal models and by more than 30% compared with open-source models. It was further used to build the DigHyd database from more than 4000 publications, containing over 30,000 entries. The workflow is also relevant to SE literature, where ionic conductivity, $E_a$, processing conditions, and stability information are often scattered across text, tables, Arrhenius plots, impedance spectra, and schematic figures.

Taken together, these studies define the extraction layer summarized in Figure 4a and illustrated by the literature-derived conductivity curation example in Figure 4b. At the same time, this layer is also where some of the main risks first appear. Extraction errors, inconsistent reporting, and literature bias can all be propagated downstream if the evidence base is treated as neutral or complete.[131,132] Recent commentary on SSE migration-barrier analysis sharpens this point further: in data-rich but methodologically inconsistent areas, LLMs are most useful as evidence-auditing tools for normalization, cohort construction, and fair comparison before physics-based validation.[120]

### 4.3 Hypothesis generation and validation-coupled reasoning

The second layer of the workflow is best understood not as open-ended "idea generation", but as validation-coupled reasoning. Once structured evidence has been assembled, the next task is to convert it into constrained, expert-facing proposals that remain anchored to downstream checks. In SE research, domain-specific assistants such as ChatSSB illustrate this route by combining literature retrieval, expert knowledge, feedback, and visualization to support question answering and research planning in solid-state battery systems.[122] More generally, reasoning systems at this layer help researchers compare material classes, organize trade-offs, and prioritize candidate directions without yet claiming autonomous decision-making.

This layer becomes more useful when the proposal is explicitly tied to validation. Uni-Electrolyte provides a representative example of an integrated AI workflow for electrolyte molecule design: beyond candidate suggestion, it combines property prediction, virtual screening, similarity search, molecular generation, retrosynthetic analysis, and reaction-network reasoning to connect molecular design with synthetic and chemical validation.[121] A more direct SE-related example is the recent framework for divalent hydride SEs, which combines a curated database, LLM-assisted insight generation, metadynamics, regression analysis, and global structure search to identify candidate compositions and evaluate them computationally.[117] Follow-up commentary on this work makes the methodological point especially clearly: the LLM does not replace first-principles modeling, but helps normalize evidence, compare materials on a fair basis, and prioritize where mechanistic validation should be applied.[120]

Beyond SE-specific systems, cross-domain database linkage can provide useful inspiration when different application fields share the same underlying chemistry.[133,134] Hydrides provide a representative example. As highlighted in the renaissance of hydrides as energy materials, hydrides are no longer limited to hydrogen storage but have become relevant to electrochemical energy storage, including SEs for mono- and divalent batteries.[135] This connection suggests that hydrogen-storage-oriented databases such as DigHyd can be linked with battery-oriented resources such as DDSE/DigBat to support hydride SE discovery.[55,136] DigHyd records may contain information on hydride stability, dehydrogenation temperature, plateau pressure, synthesis route, phase formation, dopants, and cycling behavior, whereas DDSE/DigBat organizes battery-relevant information such as ionic conductivity, activation energy, conducting ion, material family, and literature source. Although hydrogen-storage descriptors are not direct substitutes for ionic conductivity or electrochemical stability, they encode thermodynamic, kinetic, and structural features relevant to hydride-electrolyte design. A validation-coupled reasoning system could therefore search DigHyd for hydrides with suitable stability, mild synthesis conditions, and chemically flexible frameworks, and then compare them with DDSE/DigBat records on ion transport, battery compatibility, and interfacial behavior. In this way, knowledge from hydrogen-storage materials would not be transferred as direct evidence, but used as a chemically grounded source of hypotheses for identifying promising hydride SEs that still require conductivity measurements, electrochemical testing, and first-principles or molecular-dynamics validation.

### 4.4 Multi-agent AI architecture and reliability boundaries

The third layer begins when proposal-oriented systems are coupled with specialized agents, external tools, and execution modules. For SEs, BatteryAgent provides an ASSB-oriented example: it organizes a team leader, material expert, cell expert, and reviewer, and couples them with tools for knowledge retrieval, experimental data extraction, cell-performance calculation, and visualization.[137] Instead of treating ASSB analysis as a single prompt, this framework decomposes the problem into material- and cell-level subproblems and reintegrates them through reviewer validation and team-level coordination. This design is consistent with broader multi-agent formulations in which an orchestrator coordinates specialist agents, memory, and feedback rather than relying on a single monolithic model.[124] Accordingly, the central issue is not the proliferation of named systems, but the emergence of execution frameworks that translate structured reasoning into tool-linked actions.

Tool-mediated execution is the critical step. Cross-domain studies such as ChemCrow[125] and Coscientist[126] show that LLM-based systems can coordinate web searches, documentation searches, code execution, and experimental automation, thereby linking reasoning to action. These systems are not SE-specific, and they should not be presented as if they were battery-domain solutions; their value here is architectural. They define the ingredients needed for an SE-oriented execution layer: specialist agents, database and document interfaces, computation or code modules, robotic execution, and validation loops. At the materials level, A-Lab further shows how robotics, *ab initio* databases, literature knowledge, ML, and active learning can be integrated into a closed-loop platform for inorganic synthesis, providing the execution example shown in Figure 4d.[123] In early 2026, MatSource Tech, a Chinese startup, established an AI Agent driven autonomous high-throughput solid-phase experimental platform targeted at the synthesis and characterization of SEs, which is capable of producing and testing over 100 samples daily (Figure 4d). For SE discovery, these studies are best read as execution prototypes rather than finished domain solutions.

Across all three layers, reliability remains the central bottleneck. Hallucinated citations or mechanism statements can distort reasoning, literature bias can overrepresent well-studied Li-based chemistries and positive results, extraction errors can propagate into downstream screening, and data scarcity remains acute for interfaces, grain boundaries, degradation pathways, and failed experiments.[127] For SE research, this means that LLM outputs should be treated as traceable, uncertainty-aware intermediates rather than authoritative conclusions. Provenance tracking, benchmark datasets, expert curation, and physics-based validation are therefore not optional add-ons, but core design requirements for any serious LLM-enabled discovery workflow. These constraints are especially important in SE research, where sparse negative results, interfacial complexity, and inconsistent reporting can make autonomous workflows appear more mature than they actually are.[138,139]

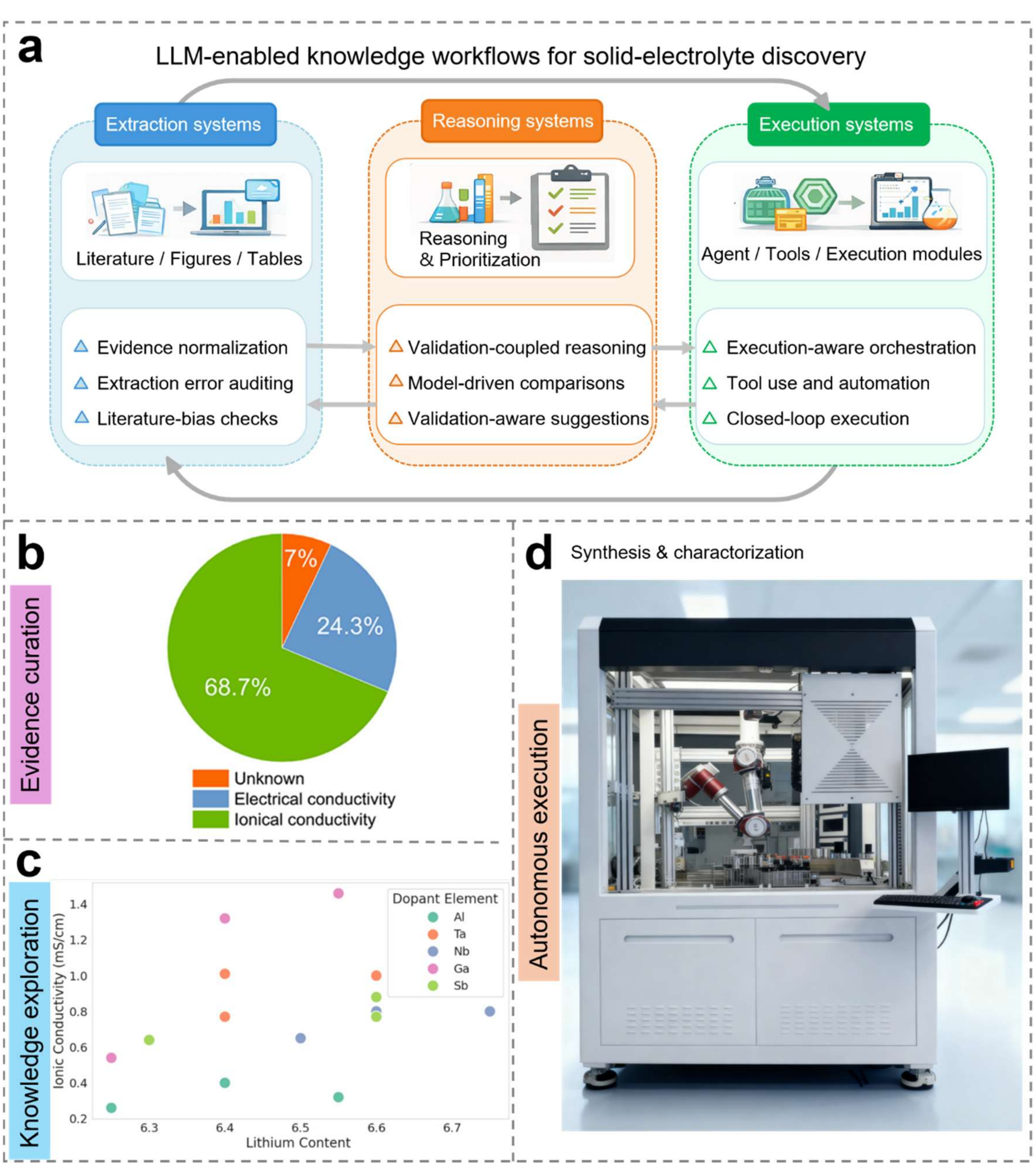

**Figure 4. LLM-enabled workflows for SE discovery.** (a) Author-designed framework linking extraction systems, reasoning systems, and execution systems through feedback. Extraction systems normalize evidence and audit literature-derived errors; reasoning systems convert structured evidence into validation-aware proposals; execution systems connect agents, tools, and computational or experimental modules. (b) Conductivity-label curation from solid-electrolyte literature, illustrating the construction of structured ionic-conductivity records.[113] (c) SE-specific knowledge exploration using literature-derived LLZO dopant–conductivity relationships.[122] (d) Robotic synthesis and characterization in an autonomous materials laboratory of MatSource Tech, illustrating how AI-guided materials selection can be connected to automated experimental execution.

## 5. Closed-Loop SE Discovery

The discovery of SEs has traditionally relied on empirical exploration and incremental compositional tuning, which becomes increasingly inefficient as the accessible chemical, structural, and processing space expands. Closed-loop, data-driven strategies provide a more systematic alternative by connecting candidate generation, simulation-based evaluation, uncertainty-aware selection, experimental validation, and model refinement into an iterative discovery cycle.[140–143] In such a workflow, ML models trained on existing experimental and computational datasets are first used to propose candidate materials with target properties, such as high ionic conductivity, wide electrochemical stability, or improved interfacial compatibility.[144] These candidates are then evaluated using first-principles calculations, MD, or MLIPs to assess phase stability, ion transport, and degradation tendencies.[145,146] Importantly, uncertainty quantification is increasingly incorporated not only to estimate the reliability of model predictions, but also to prioritize candidates that are expected to provide the greatest informational gain.[147–149]

## 5.1 From screening to feedback-driven discovery

The defining feature of closed-loop discovery is not high-throughput screening alone, but feedback. Experimental validation of selected candidates not only tests model predictions but also generates new data that can be reintegrated into the training set, thereby refining the model and guiding the next round of exploration.[150] This iterative process is particularly valuable for SEs because the design space is high-dimensional and strongly constrained by coupled requirements in transport, stability, processability, and interfaces. A material predicted to have high bulk ionic conductivity may still fail because of poor synthesizability, unstable interfaces, limited solid-solution range, or unfavorable processing behavior. Therefore, closed-loop SE discovery should be viewed as a framework that informs and progressively automates decision-making, rather than merely as a faster screening pipeline. In current practice, it helps researchers prioritize candidates, select validation experiments, and update models from feedback; in more advanced autonomous laboratories, these decisions can be executed with minimal human intervention under predefined scientific and safety constraints.

An early SE-specific example of AI-guided candidate prioritization was reported by Sendek et al., who developed an ML model to identify promising crystalline Li-ion conductors from more than 12,000 Li-containing compounds.[151] The model used structural descriptors to estimate the likelihood of fast Li-ion conduction, reducing the candidate space to a small subset for subsequent DFT-MD evaluation. This workflow identified several potential fast Li-ion conductors while also demonstrating that ML-guided selection can outperform random screening and expert intuition in the early stage of materials prioritization. Although this study did not constitute a fully autonomous closed loop, it established an important precursor to modern SE discovery workflows: using data-driven models to reduce chemical search space before applying expensive physics-based validation.

Descriptor-based learning has also been applied to the rational design of Li-argyrodite SEs. Zhao et al. developed a quantitative activation-energy prediction model and showed that anion substitution can reduce $E_a$ by balancing lattice expansion and bottleneck size under framework disorder.[152] Cation substitution can further lower $E_a$ by inducing anion disorder, while higher $Li^+$ content may activate concerted ion transport. These results highlight the value of interpretable descriptors in linking composition, local structure, and transport kinetics. At the same time, the study also illustrates an important limitation of purely predictive screening: promising compositions may not be experimentally accessible because of solid-solution limits or synthesis constraints. Incorporating synthesizability, solubility, and phase-formation likelihood into future models will therefore be essential for converting computational predictions into experimentally actionable SE candidates.

### 5.2 ML/HPC-guided discovery of halide SEs

A representative example of ML/HPC-guided SE discovery is the large-scale computational workflow reported by Chen et al. (Figure 5a-c).[153] By combining ML models with physics-based calculations on cloud high-performance computing (HPC) resources, this workflow screened more than 32 million candidates and predicted approximately half a million potentially stable materials. As summarized in Figure 5a, the candidate pool was narrowed through a funnel architecture that sequentially considered Li content, electronic insulation, electrochemical stability, thermodynamic stability, Li diffusivity, cost, mechanical softness, density, and novelty. Li self-diffusivities at 800 and 1200 K were then compared for candidates passing the earlier filters (Figure 5b), allowing materials with relatively low Li migration barriers to be prioritized.

This screening strategy identified several rare-earth halide candidates, including $Li_5YCl_8$, $Li_7Y_2Cl_{13}$, $Na_2LiYCl_6$, and a $Pna2_1$ polymorph of $Li_3YCl_6$. The $Na_xLi_{3-x}YCl_6$ ($0 \leq x \leq 3$) (NLYC) series was subsequently synthesized and experimentally characterized, revealing how Na/Li composition affects crystal structure, ionic conductivity, and activation energy (Figure 5c). This study demonstrates the power of integrating ML-based screening, physics-based filtering, HPC-scale computation, and experimental validation. At the same time, it remains closer to sequential high-throughput screening followed by targeted validation than to a fully autonomous closed-loop system. A key next step is therefore to use experimental outcomes not only as end-point confirmation, but also as feedback to update models, refine uncertainty estimates, and guide the next discovery cycle.

### 5.3 LLM-guided mining of MOF SEs

LLMs provide another route toward closed-loop SE discovery by transforming unstructured literature into searchable and model-usable knowledge. Zhang et al. used LLM-assisted literature mining and representation clustering to identify MOF SE candidates from MOF datasets and existing SE literature (Figure 5d-h).[48] In this workflow, LLM-based extraction was used to construct an MOF SE dataset, while candidate MOFs were compared with known MOF SEs through structural and chemical descriptors. A rule-based strategy further converted ligand functional groups into usable descriptors, enabling comparison between candidate MOFs and reported MOF SEs (Figure 5e). Among the screened candidates, NOTT-400 was consistently co-clustered with studied MOF SEs and was therefore selected for experimental validation.

Experimental results confirmed the feasibility of NOTT-400 as an SE. Its crystal structure is shown in Figure 5f, and TEM revealed flower-like particle assemblies formed from rod-like building units (Figure 5g). Electrochemical impedance spectroscopy demonstrated an RT ionic conductivity of $3.1\times10^{-4}$ S $cm^{-1}$, comparable to previously investigated MOF SEs within the same cluster (Figure 5h). This study is significant because it demonstrates how LLMs can contribute to SE discovery not by replacing physical models, but by organizing dispersed literature knowledge and enabling representation-based search in underexplored materials spaces. As MOF and SE databases continue to expand, similar workflows could identify additional high-potential candidates that are difficult to locate through manual literature screening.

Taken together, these studies indicate that closed-loop SE discovery is evolving along three complementary directions: descriptor-based optimization, large-scale ML-physics screening, and LLM-assisted knowledge mining. Each direction addresses a different bottleneck. Descriptor-based models improve interpretability and provide mechanistic links between composition and transport. ML-physics workflows expand the searchable chemical space while retaining physical filters. LLM-assisted approaches reduce knowledge fragmentation and enable discovery in literature-rich but structurally diverse materials classes. The central challenge is to merge these directions into a single adaptive workflow in which candidate generation, physical evaluation, experimental validation, and data curation continuously inform one another.

Overall, closed-loop methodologies mark a transition from empirical optimization toward predictive, data-centric, and self-improving SE discovery. Their future impact will depend less on the speed of screening alone than on the quality of feedback, the reliability of input data, and the inclusion of practical constraints such as synthesizability, processing compatibility, interfacial stability, and reproducibility. Continued progress in uncertainty-aware learning, multiscale simulation, high-throughput experimentation, and autonomous data capture is expected to accelerate the rational discovery of next-generation SEs.

ML/HPC-guided discovery of inorganic SEs

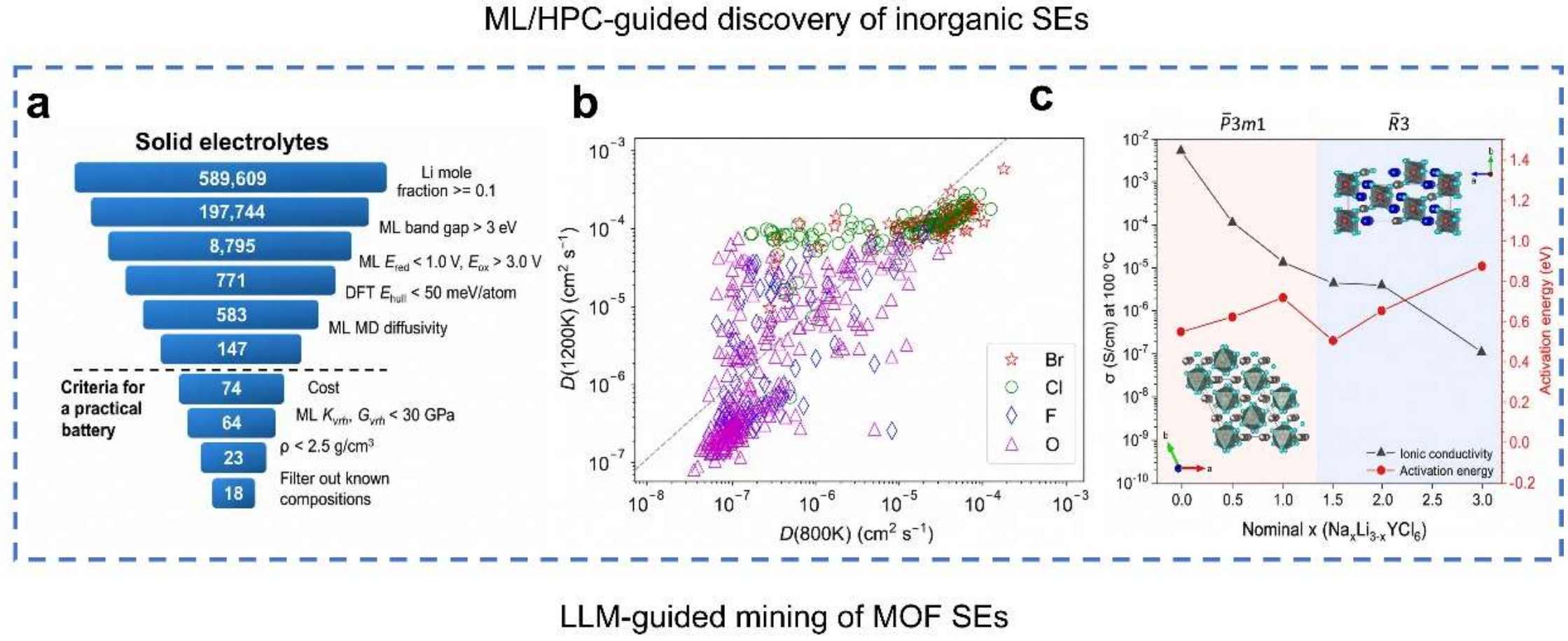

LLM-guided mining of MOF SEs

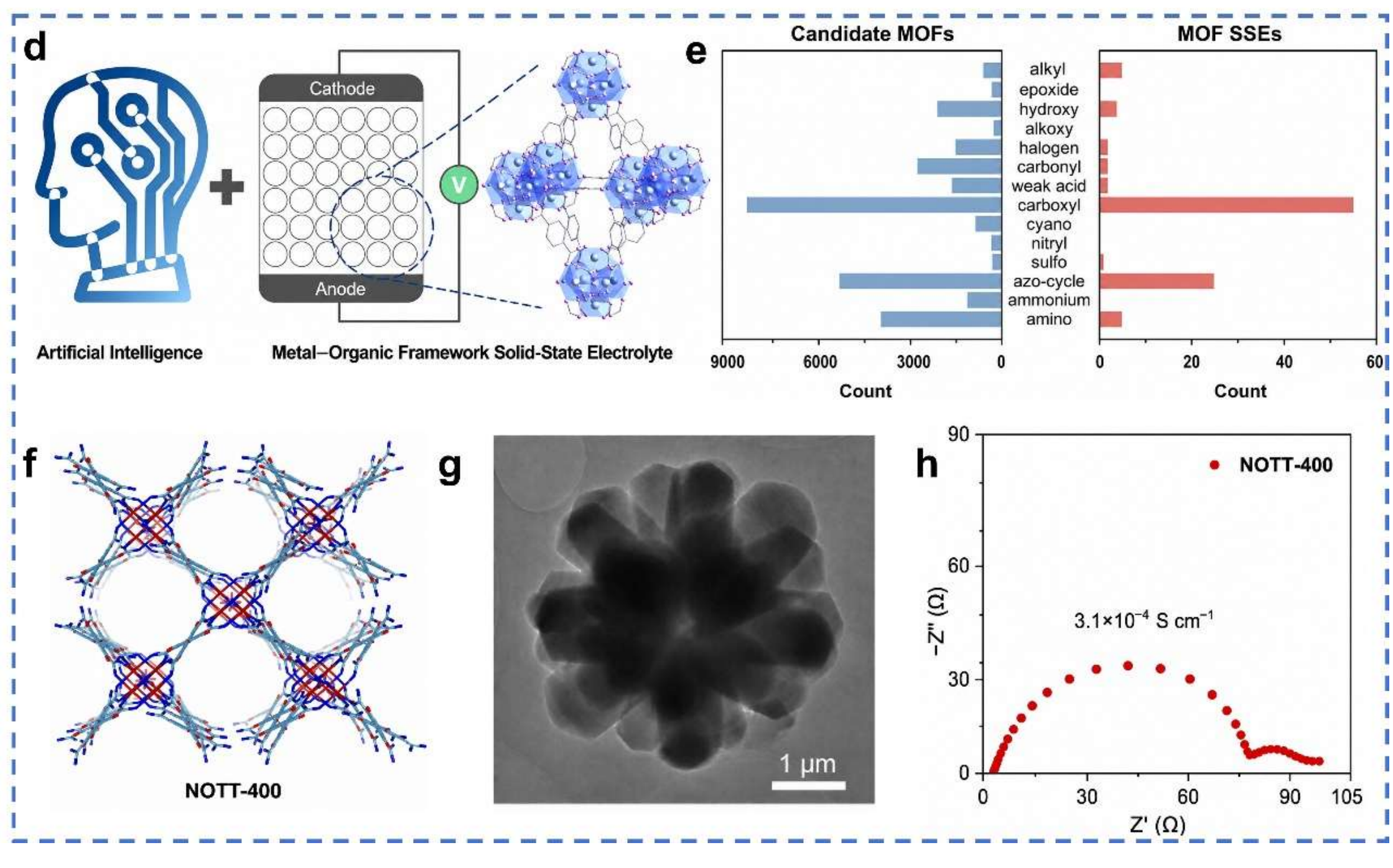

**Figure 5. Closed-loop intelligent SE discovery.** Upper row: ML/HPC-guided discovery of inorganic SEs. (a) Large-scale screening funnel used to narrow the candidate space by stability, electronic, electrochemical, diffusivity, cost, mechanical, and density filters. (b) Diffusivity-based screening of 583 candidates using Li self-diffusivity at 800 and 1200 K, grouped by anion type. (c) Experimental validation of the NLYC series, showing composition-dependent ionic conductivity at 100 °C and activation energy. (d) AI-assisted identification of MOF SE candidates from literature and database-derived information. (e) Comparison of ligand functional groups in candidate MOFs and known MOF SEs. (f) Crystal structure of NOTT-400. (g) TEM image of NOTT-400. (h) Nyquist plot showing room-temperature ionic conductivity of NOTT-400. Panels (a-c) reproduced/adapted from Ref. [153]; panels (d-h) reproduced/adapted from Ref. [48].

### 5.4 AI-assisted 3D printing for SE manufacturing

AI-assisted manufacturing is an important extension of closed-loop SE research, particularly through 3D printing. Unlike conventional powder pressing, tape casting, or slurry coating, 3D printing can encode geometry, porosity, interfacial area, and compositional gradients directly into SE architectures. This is especially relevant for ceramic oxide SEs, where ionic transport, mechanical integrity, and electrode–electrolyte contact are strongly coupled to processing history. Representative studies have demonstrated 3D-printed LLZO, LAGP, and hybrid polymer-ceramic SEs with grid corrugated, scaffold, or bicontinuous structures, showing how 3D printing can improve interfacial contact, reduce transport tortuosity, and enable mechanically compliant architectures.[9] SE printing is intrinsically data-intensive. Printable SE inks must balance SE loading, viscosity, shear-thinning behavior, curing or sintering compatibility, phase stability, and ionic conductivity. These coupled constraints make 3D printing a natural target for AI-assisted optimization, including ML models, Bayesian optimization, and generative design for identifying printable formulation windows and process–structure–property relationships. At the manufacturing and architecture level, topology optimization (TO), which optimizes material layout or geometry under physical constraints,[154] could further support the design of printable SE-containing architectures.

## 6. Outlook: Toward autonomous SE laboratories

The emergence of AI, autonomous experimentation, and data-centric materials science is rapidly reshaping the discovery paradigm for SEs. Conventional SE research has long relied on labor-intensive trial-and-error workflows involving iterative synthesis, characterization, and electrochemical testing. Although this approach has led to remarkable advances in oxide, sulfide, halide, polymer, and hydride electrolytes, the pace of discovery remains constrained by the enormous compositional space and the complexity of ion transport mechanisms. Future progress will increasingly depend on whether SE research can evolve from isolated experimental studies toward integrated autonomous laboratories in which databases, AI models, robotics, simulations, and real-time characterization are interconnected through closed-loop workflows. Such systems could continuously learn from both successful and failed experiments, enabling accelerated optimization and more reliable materials discovery.

### 6.1 Standardization and FAIR data

A foundational requirement for this transition is the establishment of standardized and FAIR SE data infrastructures. As discussed in recent research pertaining to data reliability in SE, ionic conductivity data are particularly vulnerable to ambiguity because transport properties are strongly dependent on temperature, measurement protocols, unit conventions, microstructure, and processing history. Even subtle inconsistencies in conductivity reporting, Arrhenius fitting, or annotation can propagate into curated databases and downstream ML workflows, introducing structured label noise and reducing the reliability of AI-driven predictions. Therefore, future SE databases must evolve beyond static collections of conductivity values toward context-preserving infrastructures that retain experimental provenance, processing parameters, measurement conditions, and uncertainty information.

Figure 6a illustrates this transition from conventional literature-centered workflows toward autonomous SE laboratories. In traditional workflows, experimental data are reported primarily for human interpretation, often with heterogeneous formats and incomplete metadata. By contrast, autonomous laboratories require machine-readable and interoperable data streams in which synthesis parameters, electrochemical measurements, structural characterization, and computational outputs are continuously integrated. Such infrastructures will require community-wide standardization of conductivity reporting, including explicit definitions of plotted quantities, unit conventions, fitting ranges, stack pressure, pellet density, electrode configuration, and measurement temperature. Importantly, future databases should preserve not only final standardized conductivity values but also the original experimental context and curation history, thereby enabling traceability and uncertainty-aware AI modeling.

### 6.2 Interface-aware AI

Another critical direction is the development of interface-aware AI frameworks. To date, most AI-assisted SE discovery efforts have focused primarily on bulk ionic conductivity and thermodynamic stability. However, practically ASSBs are governed not only by bulk transport but also by highly interrelated interfacial phenomena involving chemical reactions, space-charge formation, mechanical degradation, dendrite propagation, interphase evolution, and contact loss. This limitation is particularly evident in oxide electrolytes such as garnet-type LLZO, where grain boundaries and Li/SE interfaces strongly influence electrochemical performance despite high bulk conductivity. Similar challenges exist in sulfide, halide, and sodium-based systems, where interfacial decomposition and chemo-mechanical instability often dominate long-term cycling behavior.

Future AI systems must therefore move beyond scalar conductivity prediction toward physically informed and interface-aware models capable of integrating chemistry, mechanics, morphology, and electrochemical evolution simultaneously. Figure 6b illustrates a possible architecture for such interface-aware AI systems, where experimental imaging, impedance spectra, spectroscopy, atomistic simulations, and electrochemical cycling data are linked within a unified multimodal framework. The integration of operando microscopy, synchrotron characterization, MLIPs, and graph-based learning could enable AI models to identify hidden correlations between interfacial structure and electrochemical degradation. In this context, interfaces should no longer be treated as secondary corrections to bulk properties, but rather as primary design dimensions for next-generation SEs.

**6.3 B Autonomous SE discovery beyond lithium**

The future of autonomous SE laboratories will also extend far beyond lithium-based systems. Although lithium-ion conductors currently dominate SE research, growing concerns regarding lithium resource availability, cost, safety, and large-scale energy storage requirements are accelerating interest in sodium-ion, potassium-ion, magnesium-ion, calcium-ion, proton, and other multivalent-ion conductors. These emerging systems introduce fundamentally different transport mechanisms, defect chemistries, and electrochemical constraints that cannot simply be extrapolated from lithium analogues. For example, sodium superionic conductors often exhibit stronger structural flexibility and distinct grain-boundary behavior, while hydride and borohydride electrolytes display highly anharmonic lattice dynamics and complex rotational disorder. Multivalent systems further introduce correlated ion migration and stronger Coulombic interactions, which remain poorly understood.

In this regard, AI-assisted discovery may become even more important for beyond-lithium systems because experimentally accessible datasets remain relatively sparse. Autonomous laboratories combining active learning, robotics, high-throughput synthesis, and uncertainty-aware AI could efficiently explore these underdeveloped chemical spaces while continuously updating predictive models in real time. Figure 6c conceptually illustrates how future autonomous platforms may integrate literature mining, database construction, generative AI, robotic synthesis, *operando* characterization, and feedback optimization into a closed-loop discovery ecosystem spanning multiple ion chemistries. Such frameworks could significantly reduce the time required to identify promising SE compositions and accelerate the transition from empirical optimization toward mechanism-informed autonomous materials discovery.

**6.4 Digital battery ecosystem**

More broadly, future SE development should be organized as a digital materials ecosystem that connects literature-derived databases, high-throughput experiments, simulations, *operando* characterization, expert knowledge, and AI agents through shared data standards and feedback loops.[155] In this ecosystem, each component would not only support downstream prediction or validation but also return new evidence, uncertainty estimates, failed trials, and mechanistic insights to a common knowledge base. This ecosystem-level view is particularly important for SEs because ionic transport, interfacial stability, processability, and cell-level performance are strongly coupled and cannot be optimized independently. The central challenge is therefore to build continuously updated infrastructures in which data generation, model learning, experimental validation, and knowledge refinement reinforce one another. Such a digital battery ecosystem would provide the foundation for moving from case-by-case discovery toward community-scale, self-improving materials innovation.

Ultimately, the long-term vision of autonomous SE laboratories is not to replace human researchers, but to augment scientific reasoning through tighter integration between experiments, data infrastructures, AI models, and physical understanding. In this ecosystem, human researchers will remain essential for defining meaningful scientific questions, setting design constraints, interpreting mechanisms, judging unexpected results, ensuring ethical and safe operation, and deciding which autonomous outputs deserve further pursuit. The future competitiveness of SE research may depend not only on discovering individual high-performance materials but also on establishing trustworthy, interoperable, and continuously learning research ecosystems capable of transforming scientific literature into active engines for autonomous discovery.

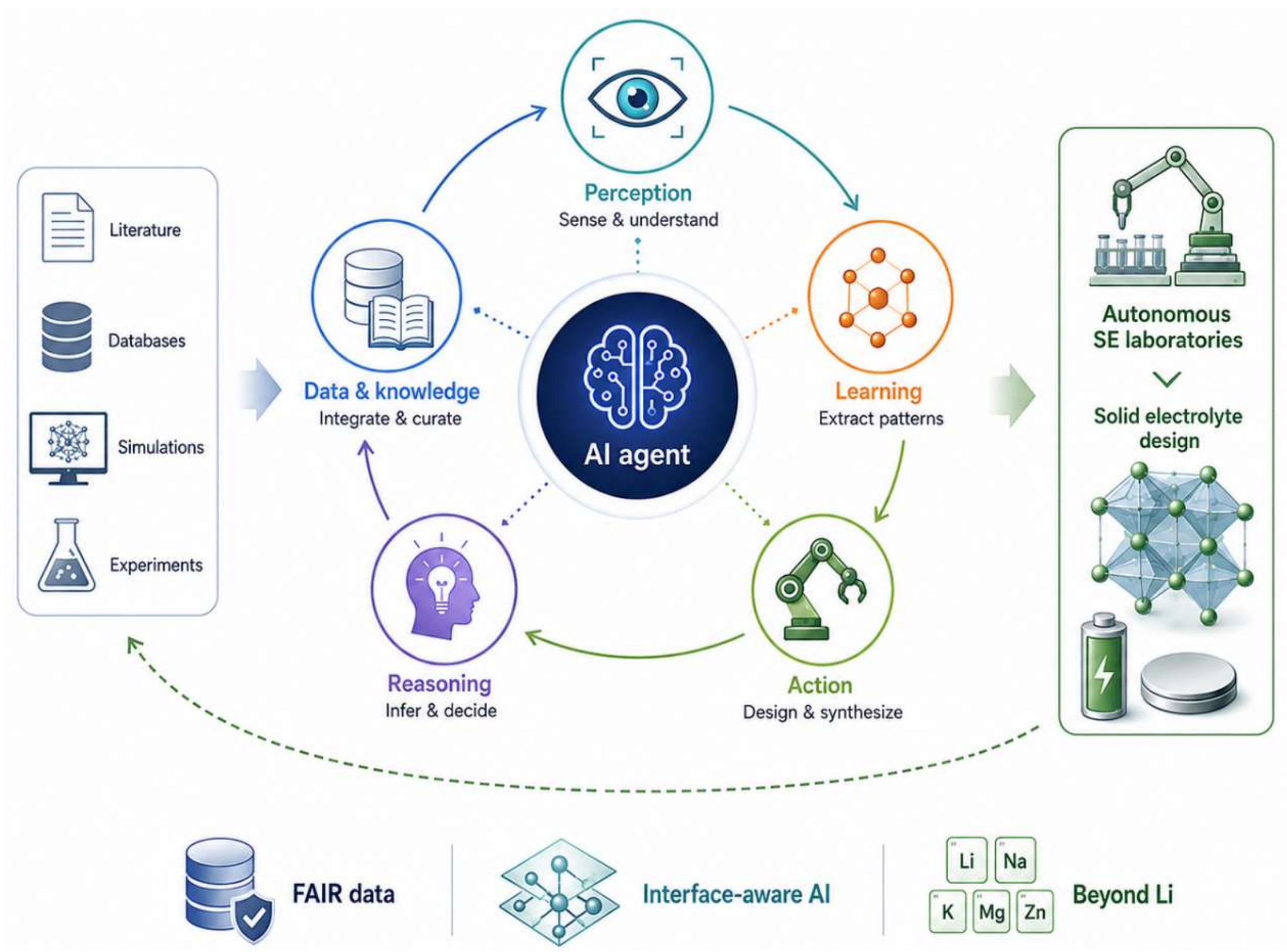

**Figure 6.** Outlook toward autonomous SE laboratories. The schematic shows a closed-loop AI-agent framework that integrates literature, databases, simulations, and experiments to guide SE design. Through perception, learning, action, and reasoning, the AI agent enables iterative optimization of composition, structure, performance, and stability, supporting standardized data use, interface-aware modeling, and discovery beyond lithium systems.

**Acknowledgements**

This work was supported by the JST Strategic International Collaborative Research Program (SICORP; Grant No. JPMJSC25E2), the Institute for Materials Research, Tohoku University (GIMRT Program, Proposal No. 202512-CRKKE-0212), and JSPS KAKENHI (Grant Nos. JP25H01508 and JP25K01737).